\documentclass[%
 reprint,
superscriptaddress,
preprintnumbers,
nofootinbib,
 amsmath,amssymb,
 aps,
 prl,
longbibliography,
twocolumn 
]{revtex4}  

\usepackage[pdftex]{graphicx}
\usepackage[utf8]{inputenc}
\usepackage{dcolumn}
\usepackage{bm}
\usepackage{amsmath}

\usepackage[colorlinks,allcolors=blue,linktocpage]{hyperref}
\usepackage{cleveref}
\crefname{equation}{Eq.}{Eqs.}%

\begin{document}
\title{Non-singular cosmology from non-supersymmetric AdS instability conjecture}
\author{Cao H. Nam}
\email{nam.caohoang@phenikaa-uni.edu.vn}  
\affiliation{Phenikaa Institute for Advanced Study and Faculty of Fundamental Sciences, Phenikaa University, Yen Nghia, Ha Dong, Hanoi 12116, Vietnam}
\date{\today}

\begin{abstract} 
We show that the non-supersymmetric AdS instability conjecture can point to how quantum gravity removes the initial Big Bang singularity, leading to a potential resolution for the past-incomplete inflationary universe. From the constraints on the dynamics of the universe realized as the nucleation of a thin-wall bubble mediating the decay of the non-supersymmetric AdS vacuum, we find the critical temperature $T_c$ and the critical scale factor $a_c$ for which the universe exists. These critical quantities are all finite and determined in terms of the parameters specifying the stringy 10D AdS vacuum solutions. Additionally, we derive the prediction of quantum gravity for $T_c$ and $a_c$ relying on the inflationary observations.
\end{abstract}

\maketitle

\emph{Introduction.}---Cosmological inflation is a rapidly accelerating expansion epoch of space in the very early universe, which can explain why the universe is big, homogeneous, isotropic, and flat as well as the absence of the magnetic monopoles \cite{Guth1981,Albrecht1982,Linde1983}. However, an inflationary universe is past-incomplete because of the presence of an initial Big Bang singularity \cite{Borde2003,Yoshida2018} where the size of the universe is zero, hence the spacetime curvature and the energy density become infinite. The Big Bang singularity is unavoidable in General Relativity (GR), which demonstrates the breakdown of GR in the extreme regime of the very early universe. On the other hand, the Big Bang singularity should be absent in a full theory of quantum gravity whose final form has not been available so far. There is another attempt based on bouncing cosmological models where the Big Bang singularity is removed via the presence of a spacetime defect \cite{Battista2021,Klinkhamer2019}.

Due to the inflation scale close to the Planck scale where the effects of quantum gravity become important, the inflation is highly sensitive to the physics in the ultraviolet (UV) regime. The effects of UV physics can break some essential futures assumed in the inflationary models. Hence, it is necessary to embed the inflation into a UV complete theory from which the inflation is obtained as an effective field theory (EFT) in the low-energy limit of the fundamental theory. In addition, this embedding opens a window to probe quantum gravity, beyond the higher-dimensional operators, which is based on the inflationary observations measured from the spectrum of cosmic microwave background (CMB) and the spectrum of primordial gravitational waves.

The recent progress exploring quantum gravity aims to determine its universal and fundamental principles usually stated as the conjectures whose establishment is mainly based on the physics of black holes, the compactification of string theory, and the general arguments of quantum gravity \cite{Palti2019}. These conjectures impose the constraints on the low-energy EFTs that are UV completed into quantum gravity. The Big Bang singularity is inconsistent with quantum gravity, thus it is reasonable to think that universal and fundamental principles of quantum gravity can tell us a mechanism in UV complete theory of gravity to create a non-singular universe whose critical scale factor and critical temperature representing the smallest size and the highest temperature for its existence are all finite.

The non-SUSY AdS instability conjecture \cite{Ooguri2017} indicates that an anti-de Sitter (AdS) vacuum in quantum gravity without being protected by supersymmetry (SUSY), or it is supported by flux, must decay as a result of applying the weak gravity conjecture \cite{Harlow2023} for $p$-forms. This conjecture has many implications for particle physics and cosmology like the connection between the neutrino masses and the observed cosmological constant \cite{Martin-Lozano2017,Hamada2017,Nam2023b,Nam2023c}. In particular, the non-SUSY AdS instability conjecture leads to a way to embed the dS universe in string theory considered as UV complete theory of quantum gravity where the dS universe emerges from the decay of the false non-SUSY AdS vacuum to the true SUSY AdS vacuum \cite{Banerjee2018}. There are obstructions to constructing dS vacuum solutions in string theory, motivating the dS conjecture \cite{Obied2018} stating that dS vacua do not exist in quantum gravity and the studies of the dS instability \cite{Krishnan2019}. The proposal in \cite{Banerjee2018} can provide a resolution for matching string theory with the observations about the late-time accelerating expansion of the universe.\footnote{An alternative to make string theory consistent with observed dS universe, based on the AdS vacuum, is to compactify the AdS$_5$ factor on a circle and take into account the nontrivial dynamics of the 4D tensor component \cite{Nam2023a}.} This cosmological scenario has been extended in the presence of a cloud of strings \cite{Koga2019} and quintessence \cite{Koga2021}, with Kerr-AdS$_5$ spacetime \cite{Koga2023}, and for the AdS compactification of M-theory over a squashed four-sphere \cite{Dibitetto2023}. The probability of bubble nucleation is connected to Vilenkin's amplitude in quantum cosmology \cite{VanRiet2021}.

In this work, we show how quantum gravity removes the Big Bang singularity via the non-SUSY AdS instability conjecture. Our universe is realized as the nucleation of a thin-wall bubble mediating the decay of the false non-SUSY AdS vacuum to the true SUSY AdS vacuum. In particular, from the constraints on the dynamics of the bubble and Heisenberg's uncertainty principle, we determine the critical temperature and the critical scale factor for the existence of the universe. These critical quantities are finite, hence the beginning of the universe avoids the Big Bang singularity. In order to determine the critical temperature and the critical scale factor in terms of the parameters of quantum gravity, we study the supersymmetry and gauge symmetry of the critical points of 5D supergravity and their uplift to the 10D AdS vacuum solutions of string theory compactified on a 5D internal manifold. Then, we express these critical quantities in terms of the units of the self-dual five-form flux stabilizing the size of the 5D internal manifold. Interestingly, as soon as the universe was created by the vacuum decay, it would enter the inflation era. Using the inflationary observations, we derive the prediction of quantum gravity for the critical temperature and the critical scale factor.

\emph{Creation of the non-singular universe.}---Let us start with two $\text{AdS}_5$ vacua denoted by ``$+$" and ``$-$" corresponding to the following element
\begin{equation}
ds^2_{5,\pm}=-(1+k^2_{\pm} r^2)d\tau^2_{\pm}+\frac{dr^2}{1+k^2_{\pm}r^2}+r^2d\Omega^2_3,\label{AdSvac-metr}  
\end{equation}
where $k_->k_+$ and $d\Omega^2_3$ is the metric of three-sphere $S^3$. The $\text{AdS}_5$ vacuum with the higher energy corresponding to the cosmological constant $\Lambda_+=-6k^2_+$ is unstable and thus it decays to the $\text{AdS}_5$ vacuum with the lower energy corresponding to the cosmological constant $\Lambda_-=-6k^2_-$. Through the Coleman-de Luccia transition \cite{Coleman1980}, the vacuum decay is mediated by the nucleation of a thin-wall bubble or spherical brane which is realized as our universe.

The metric that is induced on the bubble or brane universe reads 
\begin{equation}
ds^2_{\text{bub.}}=-dt^2+\frac{a(t)^2}{\mathcal{K}}d\Omega^2_3,    
\end{equation}
where $a(t)$ is the scale factor related to the radial coordinate $r$ of the AdS bulk as $r=a(t)/\sqrt{\mathcal{K}}$ and $\mathcal{K}>0$ is the curvature of the spatial three-sphere at the present-day time $t_0$ with $a(t_0)=1$. The dynamics of the universe is governed by the second Israel junction condition \cite{Delsate2014} as follows
\begin{eqnarray}
M^3_-\left[K^{(-)}_{\mu\nu}-h_{\mu\nu}K^{(-)}\right]-M^3_+\left[K^{(+)}_{\mu\nu}-h_{\mu\nu}K^{(+)}\right]=S_{\mu\nu},\nonumber\\
\end{eqnarray}
where $K^{(\pm)}_{\mu\nu}=e^M_\mu e^N_\nu\nabla_Mn^{(\pm)}_N$ is the extrinsic curvature of the brane [with $e^M_\mu$ and $n^{(\pm)}_N=(-\dot{a}(t)/\sqrt{\mathcal{K}},\dot{\tau}_{\pm}(t),0,0,0)$ are the tangent vector and the unit normal vector to the brane, respectively], $K^{(\pm)}=h^{\mu\nu}K^{(\pm)}_{\mu\nu}$, $S_{\mu\nu}$ is the energy-momentum tensor on the brane, and $M_\pm$ are the 5D Planck scales on two sides of the brane. It should be noted here that the 5D Planck scale generally differs on either side of the brane. However, in order to show the present scenario simply without loss of generality, we will assume $M_-=M_+\equiv M_5$ in the following calculations. From this equation, we find
\begin{eqnarray}
 \left(\frac{\dot{a}}{a}\right)^2+\frac{\mathcal{K}}{a^2}&=&\frac{\left(k^2_1-\hat{\rho}^2\right)\left(k^2_2-\hat{\rho}^2\right)}{4\hat{\rho}^2},\label{Fri-equ}
\end{eqnarray}
where the dot denotes the derivatives with respect to $t$, $k_{1,2}\equiv k_-\pm k_+$, and $\hat{\rho}\equiv(\sigma+\rho_r)/(3M^3_5)$ where $\sigma$ and $\rho_r$ refer to the brane tension and the energy density of the radiation, respectively, which besides the curvature are dominant components over all other species at the very early universe. The usual Friedmann equation can be restored in the regime of $\mathcal{\sqrt{K}}/a,\dot{a}/a\ll k_{\pm}$ corresponding to the fact that the reduced brane tension $\hat{\sigma}\equiv\sigma/(3M^3_5)$ is very close to $k_2$, where we can identify the observed Planck scale as 
\begin{eqnarray}
M^2_{\text{Pl}}=\frac{k_--k_+}{2k_-k_+}M^3_5.\label{4DMPl}   
\end{eqnarray}

We observe from Eq. (\ref{Fri-equ}) that quantum gravity determines an initial condition for the beginning of the universe in which the universe has a positive spatial curvature corresponding to a closed universe. This is consistent with the Planck 2018 data that without the lensing likelihood prefers to the universe of the positive spatial curvature at a ratio of $50 : 1$ against a spatially flat universe \cite{Planck2018}. The presence of the positive spatial curvature can resolve the tensions between the CMB data with other measurements like the CMB lensing, the baryon acoustic oscillations, and Supernovae Ia \cite{Valentino2019,Handley2021}. (There have been attempts to combine the Planck data with reliable external datasets to break the geometrical degeneracy, which is towards a spatially flat universe \cite{Vagnozzi2021,Gariazzo2021,Dhawan2021}.) With the best-fit value of the present-day curvature density $\Omega_{\mathcal{K},0}\equiv-\mathcal{K}/(a_0H_0)^2=-0.0438$ and the present-day Hubble parameter $H_0=67.4$ km s$^{-1}$ Mpc$^{-1}$ \cite{Planck2018}, we can find $\mathcal{K}\approx2.2\times10^{-9}$ Mpc$^{-2}$ and the present-day curvature radius about $2.1\times10^4$ Mpc.

For a non-singular universe, there is a critical temperature $T_c$ and a critical scale factor $a_c$, which are all finite, representing the highest temperature and the smallest (rescaled) size of the universe, respectively. We shall show that the decay of the unstable AdS vacuum to the stable AdS vacuum would create such a universe, allowing the universe to avoid the initial Big Bang singularity. Because the energy density of the radiation on the brane universe behaves in the scale factor as $\rho_r\propto a^{-4}$ and the temperature as $\rho_r\propto T^4$, $\rho_r$ and $T$ would grow when decreasing the size of the universe. However, in the present cosmological scenario, the size of the universe cannot be arbitrarily small or the universe temperature would never diverge because there is an upper bound for the energy density of the radiation on the brane. 

\begin{figure}[t]
 \centering
\begin{tabular}{cc}
\includegraphics[width=0.35 \textwidth]{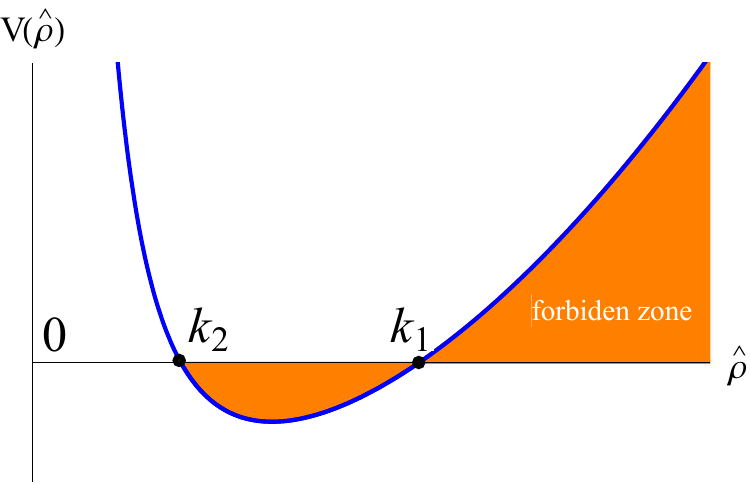}
\end{tabular}
 \caption{The behavior of $V(\hat{\rho})$ that denotes the right-hand side of Eq. (\ref{Fri-equ}) in terms of the reduced energy density $\hat{\rho}$ on the brane.}\label{pot-rho}
\end{figure}

For the decay of the unstable AdS vacuum through the bubble nucleation happening, the energy density on the brane should not exceed the critical value $3M^3_5(k_--k_+)$, as depicted in Fig. \ref{pot-rho}.
In addition, in the situation that the energy density of the radiation on the brane grows due to decreasing the size of the universe, the brane universe cannot transfer the energy-momentum from the brane to the bulk to guarantee the aforementioned upper bound. This can be realized from the conservation law along the brane given by $\nabla^\mu S_{\mu\nu}=R_{MN}n^Me^N_\nu$ where $R_{MN}$ is the Ricci tensor of the AdS bulk \cite{Clifton2011}. The term $R_{MN}n^Me^N_\nu$ which describes the interchange of the energy-momentum between the brane and the AdS bulk vanishes identically with the bulk geometry given by Eq. (\ref{AdSvac-metr}). Hence, there is no exchange of the energy-momentum between the brane and the AdS bulk. As a result, the energy density of the radiation on the brane needs to satisfy the following condition $ \rho_r\lesssim6k_-k_+M^2_{\text{Pl}}-\sigma$ where we have used Eq. (\ref{4DMPl}). Then, we find the critical temperature for which the brane universe exists as
\begin{eqnarray}
T_c\simeq\left[\frac{30\left(6k_-k_+M^2_{\text{Pl}}-\sigma\right)}{\pi^2g_*}\right]^{1/4},
\end{eqnarray}
where $g_*$ is the relativistic number of degrees of freedom near the critical temperature. The critical scale factor $a_c$ can be estimated from Heisenberg's uncertainty principle where the momentum of a relativistic particle is given by $T_c$ and the size of the volume containing the particle is about $2\pi a_c/\sqrt{\mathcal{K}}$. This leads to
\begin{eqnarray}
a_c\simeq\frac{1}{2\pi}\left[\frac{\pi^2g_*\mathcal{K}^2}{30\left(6k_-k_+M^2_{\text{Pl}}-\sigma\right)}\right]^{1/4}.
\end{eqnarray}

\emph{$T_c$ and $a_c$ in UV complete theory of gravity.}---Because the non-singular universe is created in the context of quantum gravity, an ultraviolet (UV) complete theory of gravity shall tell us in a more precise way about the critical temperature and the critical scale factor for the existence of the universe.

The non-SUSY/SUSY AdS$_5$ vacua correspond to critical points of the supergravity potential of 5D supergravity $V(\alpha,\chi)$ given by \cite{Pilch2000}
\begin{eqnarray}
V(\alpha,\chi)=\frac{g^2}{3}\left(\frac{3}8\left|\partial_\chi W\right|^2+\frac{1}{16}\left|\partial_\alpha W\right|^2-\left|W\right|^2\right), 
\end{eqnarray}
where $\alpha$ and $\chi$ are the supergravity fields, $g$ is the gauge coupling, and $W$ is the superpotential given as follows
\begin{eqnarray}
W(\alpha,\chi)=\frac{1}{4e^{2\alpha}}\left[\cosh(2\chi)(e^{6\alpha}-2)-(3e^{6\alpha}+2)\right]. 
\end{eqnarray}
All critical points of the supergravity potential and their supersymmetry and gauge symmetry are analyzed in \cite{Gunaydin1985,Gunaydin1986,Khavaev2000}. The potential at these critical points is negative, they thus admit the AdS metrics. Because the 5D supergravity is a consistent truncation of the 10D IIB supergravity, we can uplift the 5D AdS vacua to the 10D AdS vacuum solutions of string theory compactified on a proper 5D internal manifold. The gauge symmetry of the critical point would be the isometry of the 5D internal manifold.

The stable SUSY AdS vacuum corresponds to a critical point at $\alpha=0$ and $\chi=0$ which preserves all supersymmetries (i.e. $\partial_iV=0$ and $\partial_iW=0$ with $i=\alpha,\chi$). This critical point leads to the SUSY AdS vacuum with the cosmological constant $\Lambda_-=-3g^2/2$. Also, this critical point has the SO(6) gauge symmetry which is the isometry of the round five-sphere $S^5$, hence this AdS vacuum can be uplifted to the compactified solution of the IIB string theory on the round five-sphere, i.e. $\text{AdS}_5\times S^5$. As a result, the parameter $k_-$ or the gauge coupling $g$ is determined by $N_5$ units of self-dual five-form flux which stabilizes $S^5$ as follows
\begin{eqnarray}
k_-=\frac{g}{2}=2M_5\left(\frac{\pi}{N_5}\right)^{2/3},\label{kmn}
\end{eqnarray}
where the 5D Planck scale $M_5$ is related to the string coupling $g_s$, the string length $l_s$, and the volume of the internal manifold $\text{Vol}_5$ as $M^3_5=(2\pi)^{-7}\text{Vol}_5/(g^2_sl^8_s)$. 

The unstable non-SUSY AdS vacuum corresponds to another critical point at $\alpha=0$ and $\chi=\text{arccosh}(2)/2$ which breaks completely supersymmetry (i.e. $\partial_iV=0$ and $\partial_iW\neq0$ with $i=\alpha,\chi$). This critical point has the SU(3)$\times$U(1) gauge symmetry which is the isometry of the squashed five-sphere which is treated as a U(1) bundle over the base $\mathbb{CP}^2$ \cite{Imamura2013}. Thus, this critical point can be uplifted to the compactified solution of the IIB string theory on the squashed five-sphere with the squashed parameter $\nu=\sqrt{2/3}$ \cite{Romans1985}. All supersymmetries are broken by the presence of the three-form flux sourced by the brane mediating the vacuum decay. In addition, the three-form flux squashes $S^5$, leading to break the SO(6) isometry to the SU(3)$\times$U(1). In general, the deformation of the internal manifold does not preserve the number of five-form flux. Then, we determine $k_+$ in terms of $N'_5$ units of five-form flux in the non-SUSY AdS vacuum as follows
\begin{eqnarray}
k_+&=&\frac{3M_5}{\sqrt{2}}\left(\frac{\pi}{2N'_5}\right)^{2/3}.\label{kpls}
\end{eqnarray}

The decay of the non-SUSY AdS vacuum to the SUSY AdS vacuum is mediated by the nucleation of a spherical $(p,q)$-five brane \cite{Banerjee2018} which is a bound state of $p$ D5-branes and $q$ NS5-branes \cite{Aharony1998} and sources the three-form flux breaking all supersymmetries and squashing $S^5$. The $(p,q)$-five brane is wrapped around $S^2$ in the 5D internal space (topologically equivalent to $S^5$), which can be stabilized using Myers's mechanism \cite{Myers1999}. $N$ D3-branes perturbed with transverse three-form fields may polarize into $M$ five-branes with the polarization potential given as \cite{Polchinski2000} $V_{\text{pol.}}=M^2R^4/N-c_3MR^3-m^2NR^2$ where $R$ is the radius of $S^2$, $c_3$ determines the field strength of three-form fields, and $m$ is a mass parameter. With $m^2>0$, the polarization potential always has a minimum, which implies a stable configuration for five-branes.

The effective tension of the BPS $(p,q)$-five brane mediating the vacuum decay can be computed based on the tension of the domain wall interpolating two vacua. It is related to the change of the superpotential across the domain wall \cite{Gukov2000}, given as
\begin{eqnarray}
\sigma_{\text{BPS}}&=&M^3_5g\left|W(0,0)-\tilde{W}\left(0,\text{arccosh}(2)/2\right)\right|\nonumber\\
&=&M^4_5\left(\frac{\pi}{N_5}\right)^{2/3}\left[6-7\left(\frac{N_5}{2N'_5}\right)^{2/3}\right],
\end{eqnarray}
where the superpotential at the non-SUSY critical point is rescaled as $\tilde{W}=2^\alpha W$
so that $V_-=\Lambda_-/2$.

\begin{figure}[t]
 \centering
\begin{tabular}{cc}
\includegraphics[width=0.4 \textwidth]{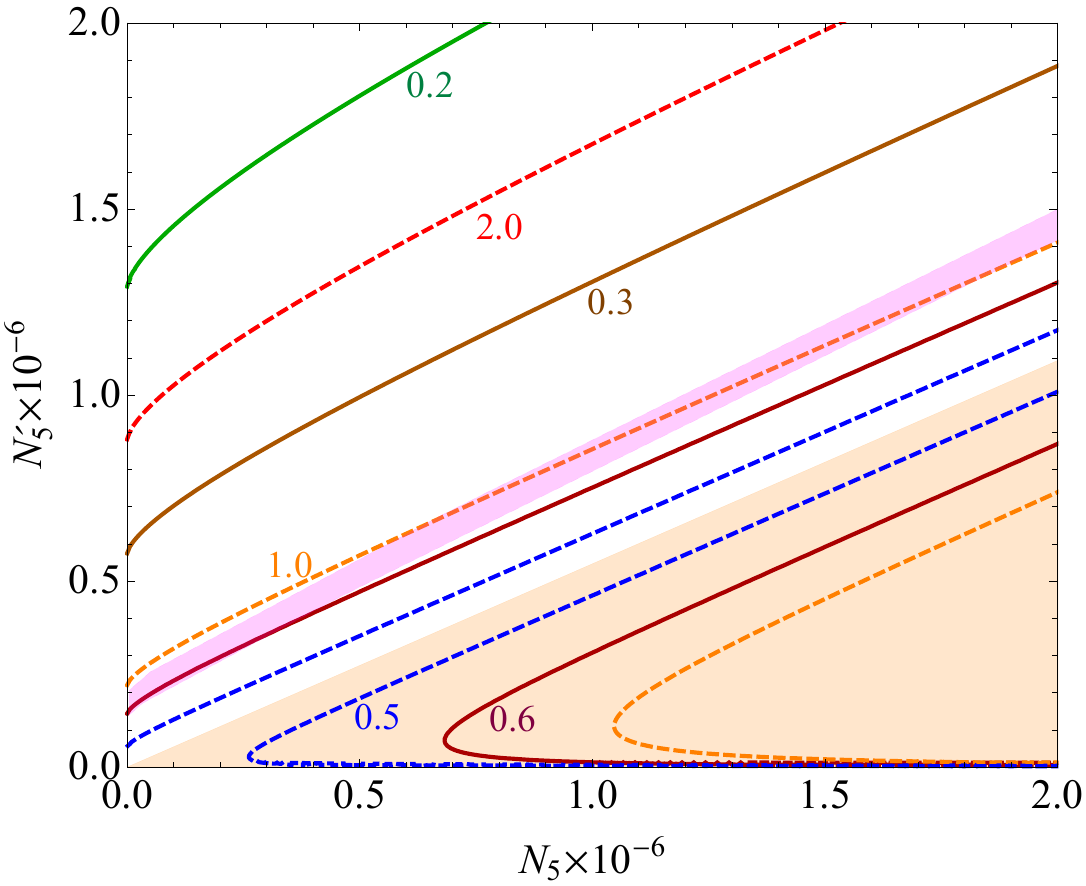}
\end{tabular}
 \caption{The contours of $T_c\times10^{-16}/\text{GeV}$ (as the solid curves) and $a_c\times10^{59}$ (as the dashed curves) in the $(N_5,N'_5)$ plane, with $g_*\simeq106.75$. The magenta and light orange regions refer to the inflation constraint and no vacuum decay, respectively.}\label{cont-Tcac}
\end{figure}

In general, the critical temperature $T_c$ and the critical scale factor $a_c$ depend on $N_5$
units of five-from flux in the SUSY AdS vacuum and the distortion of the internal manifold that can change the number of five-form flux in the non-SUSY AdS vacuum. In Fig. \ref{cont-Tcac}, we show the values of $T_c$ (as the solid curves) and those of $a_c$ (as the dashed curves) in terms of $N_5$ and $N'_5$ with $g_*\simeq106.75$ (the relativistic number of degrees of freedom for the Standard Model).
The light orange region does not lead to the decay of the non-SUSY AdS vacuum to the SUSY AdS vacuum due to $k_-<k_+$ yielding $N'_5/N_5<3\sqrt{3}/2^{13/4}$. The allowed values of $T_c$ and $a_c$ belong to the white region. When incorporating inflation, there is an additional constraint shown by the magenta band, which we shall study below. In the case that the number of five-form flux is untouched in the vacuum decay, the critical temperature $T_c$ and the critical scale factor $a_c$ are completely determined by a unique quantity in the UV complete theory as follows
\begin{eqnarray}
T_c\sim\frac{M_{\text{Pl}}}{N^{1/2}_5},\ \ a_c\sim N^{1/2}_5\left(\frac{\mathcal{K}}{M^2_{\text{Pl}}}\right)^{1/2}.
\end{eqnarray}

\emph{Constraint from inflation.}---The energy density of the radiation decreases with the expansion of the universe leading to the right-hand side of Eq. (\ref{Fri-equ}) being nearly constant given by $(k^2_1-\hat{\sigma}^2)(k^2_2-\hat{\sigma}^2)/(4\hat{\sigma}^2)$. This results in a phase of inflation shortly after the universe was created by the decay of the false AdS vacuum to the true AdS vacuum. 

Inflation ends and is followed by the reheating as an energy part of the false non-SUSY AdS vacuum is converted to the matter in the bulk. Inflation stops when the energy of the false non-SUSY AdS vacuum decreases from the initial value ($\propto-k^2_+$) to a lower one such that the right-hand side of Eq. (\ref{Fri-equ}) is near zero. This means that after the end of the inflation, the spherical $(p,q)$ five-brane continues to expand due to the vacuum decay, but at a much slower rate.

The end of the inflation requires the backreaction of additional fields in the exterior region of the AdS bulk. For instance, we consider the backreaction of the dilaton evolution on the geometry of the non-SUSY AdS vacuum. The energy density of the non-SUSY AdS vacuum $\sim-M^3_5k^2_+$ can be expressed by a function $\propto-g^{-5/4}_s$ where $g_s$ is determined by the dilaton $\phi$ as $g_s=e^{\phi}$. This implies that the dilaton evolution in the false vacuum would lead to a change in the energy density of this vacuum. [As the dilaton changes its values, it would lead to the change of the 5D Planck scale in the non-supersymmetric AdS vacuum. For this 5D Planck scale to be kept fixed, $N'_5$ should be rescaled corresponding to taking into account the backreaction of the moduli dynamics.] During the inflation, the dilaton evolution is slow in such a way that the value of dilaton approximately remains constant during this stage. The inflation ends as the dilaton approaches its minimum. In addition, the presence of additional fields in the exterior region of the AdS bulk contributes an ADM mass term in the line element of the false AdS vacuum which plays the role of the radiation with respect to the observer on the bubble universe \cite{Kraus1999,Danielsson2019}. This can allow us to realize the reheating after inflation.
 
In this way, the backreaction of additional fields on the false non-SUSY AdS vacuum would modify the metric outside the bubble. As a result, in Eq. (\ref{Fri-equ}), we need to modify $k_+$ in a general way as follows
\begin{eqnarray}
k^2_+&\longrightarrow& k^2_++U_0\xi(a),
\end{eqnarray}
where $U_0$ satisfies $k_--\sqrt{k^2_++U_0}\simeq\sigma$ such that the inflation ends and $\xi(a)$ is a function of the scale factor. When the backreaction is small as seen by an observer on the bubble universe, corresponding to the inflation era, the function $\xi(a)$ is close to zero. However, when the backreaction becomes significant, the inflation ends and it is followed subsequently by the reheating, which implies the behavior of the $\xi(a)$ as $\xi(a)\simeq 1-\rho_{\text{re}}a^{-4}$ where $\rho_{\text{re}}$ is the energy density of the radiation at the reheating. Phenomenologically, we simply adopt $\xi(a)$ as follows
\begin{eqnarray}
\xi(a/a_e)=1-q\times\left(\frac{a_e}{a}\right)^{4[\omega(a/a_e)+1]},\label{xifun}
\end{eqnarray}
where $q$ is constant, $a_e$ is the scale factor at the end of inflation, and $\omega$ is a function of the scalar factor. During the inflation, $q(a_e/a)^{4(\omega+1)}$ is close to one, and thus $\omega$ is near to $-1$. The inflation stops when $\omega$ changes drastically from $-1$ to zero around $a_e$. 

With $\Omega_{\mathcal{K},0}=-0.0438$ and the pivot scale given by $k_*=a_*H_*=0.05$ Mpc$^{-1}$ \cite{Planck2018}, we can calculate the curvature density parameter at the time which $k_*$ crosses the comoving Hubble horizon as $\Omega_{\mathcal{K},*}=8.84\times10^{-7}$. This implies that the mode $k_*$ is unaffected by the non-zero spatial curvature and hence we can calculate primordial cosmological parameters using the flat approximation. 

\begin{figure}[t]
 \centering
\begin{tabular}{cc}
\includegraphics[width=0.4 \textwidth]{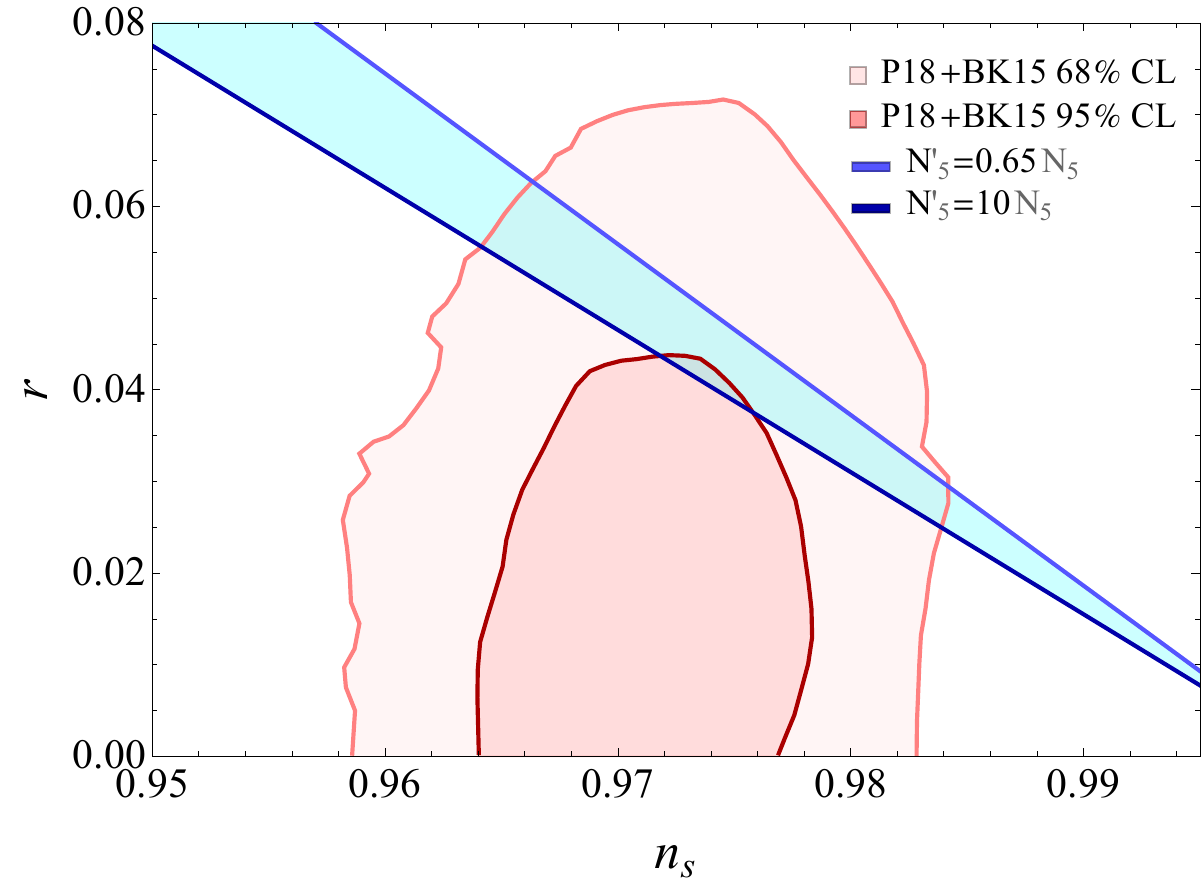}
\end{tabular}
 \caption{The tensor-to-scalar ratio versus the scalar spectral index, predicted in the present cosmological scenario represented by the blue band.}\label{ns-r-plot}
\end{figure}

The scalar spectrum index $n_s$ and the tensor-to-scalar ratio $r$ are determined as $n_s=1-4\epsilon_*+2\eta_*$ and $r=16\epsilon_*$ where $\epsilon_*$ and $\eta_*$ are the values of two Hubble slow-roll parameters defined by $\epsilon=-\dot{H}/H^2$ and $\eta=\ddot{H}/(\dot{H}H)$ at the horizon crossing time (denoted by $*$). They are given by
\begin{eqnarray}
\epsilon_*&\simeq&\frac{4U_0(k_1k_2+\hat{\sigma}^2)}{(k^2_1-\hat{\sigma}^2)(k^2_2-\hat{\sigma}^2)}(1+\omega_{*})qe^{4N_*(\omega_{*}+1)},\nonumber\\
\eta_*&\simeq&-\epsilon_*-4(1+\omega_{*}),
\end{eqnarray}
where $N_*$ refers to the number of e-folds after the mode $k_*$ crosses the comoving Hubble horizon. Note that, $qe^{4N_*(\omega_{*}+1)}$ is approximately one during inflation. We represent the inflation prediction due to the decay of the non-SUSY AdS space with an additional source described by the function $\xi(a)$ given by Eq. (\ref{xifun}) in Fig. \ref{ns-r-plot}, expressed by the blue band.
In addition, we show the $1\sigma$ and $2\sigma$ confidence contours for $(n_s,r)$ from the PLANCK 2018 \cite{Planck2019} and BICEP2 \cite{BICEP2018} with an extension of the positive spatial curvature for $\Lambda$CDM model \cite{Hergt2022}. We observe that the case of $N'_5<N_5$ is less to favor the data compared to the case of $N'_5>N_5$. 

\begin{figure}[t]
 \centering
\begin{tabular}{cc}
\includegraphics[width=0.45 \textwidth]{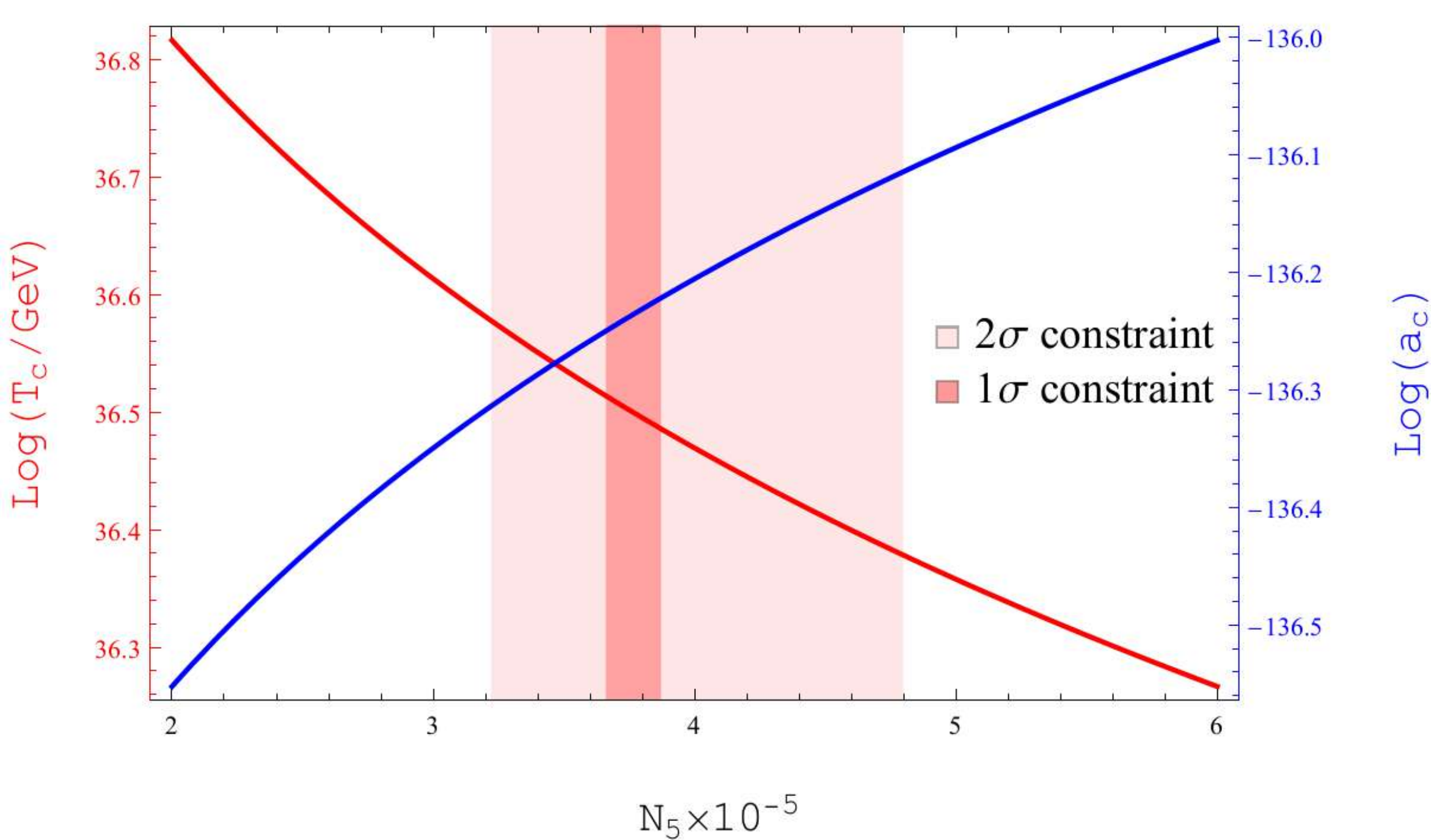}
\end{tabular}
 \caption{The prediction of $T_c$ and $a_c$ for the case that the number of five-form flux is untouched, which agrees with the $1\sigma$ and $2\sigma$ contours of $(n_s,r)$.}\label{Tc-ac-N5-plot}
\end{figure}

The scalar power spectrum amplitude $A_s\approx2.1\times10^{-9}$ \cite{Planck2019} is calculated in terms of the Hubble parameter $H_i$ during inflation and the tensor-to-scalar ratio $r$ as $A_s=2H_i^2/(\pi^2M^2_{\text{Pl}}r)$. Using this relation and the inflation constraint just studied above, we can derive the prediction of quantum gravity for the critical temperature and critical scale factor. The prediction is represented by the magenta band given in Fig. \ref{cont-Tcac}. In the case of $N'_5=N_5$, we show more explicitly the prediction in Fig. \ref{Tc-ac-N5-plot}. 
Accordingly, the critical temperature and the critical scale factor of the universe are estimated as follows
\begin{eqnarray}
T_c\sim10^{16}\ \text{GeV},\ \ a_c\sim10^{-60}.  
\end{eqnarray}
This critical value is in the order of the grand unification scale. In addition, the critical radius of the universe is above the Planck length scale, hence the universe can be quite well described by GR.

With $N_5$ and $N'_5$ being of order $10^5$, we need a large number of five-branes to generate the three-form flux breaking SUSY. This implies a large number of massless modes corresponding to the non-Abelian gauge fields on a stack of five-branes. The corresponding gauge coupling is given by $g^2_{\text{YM}}=g_sl^2_s/\text{Vol}(S^2)$ where  $\text{Vol}(S^2)$ is the volume of $2$-sphere on which the five-branes wrap. This coupling is strongly suppressed by the string scale $l_s$ and as a result, these non-Abelian gauge fields would be decoupled in the low-energy regime. In addition, the $(p,q)$-five brane that mediates the vacuum decay is the BPS one which is a stable configuration. Thus, there are no tachyonic modes living on the worldvolume of the $(p,q)$-five brane.

I thank N. A. Ky, N. T. H. Van, P. V. Ky, and T. N. Hung for their comments and discussions. I would like to express sincere gratitude to the Referee for his/her constructive comments,
suggestions and questions by which the quality of the paper has been improved.

\end{document}